RESEARCH ARTICLE

# Deciphering Interphase Instability of Lithium Metal Batteries with Localized High-Concentration Electrolytes at Elevated Temperatures

Tao Meng, Shanshan Yang, Yitong Peng, Xiwei Lan, Pingan Li, Kangjia Hu, and Xianluo Hu*

[*] T. Meng, S. S. Yang, Y. T. Peng, X. W. Lan, P. A. Li, K. J. Hu, Prof. X. L. Hu
State Key Laboratory of Materials Processing and Die & Mould Technology
School of Materials Science and Engineering
Huazhong University of Science and Technology
Wuhan 430074, China
E-mail: huxl@mail.hust.edu.cn (X. L. Hu)

**Abstract:** Lithium metal batteries (LMBs), when coupled with a localized high-concentration electrolyte and a high-voltage nickel-rich cathode, offer a solution to the increasing demand for high energy density and long cycle life. However, the aggressive electrode chemistry poses safety risks to LMBs at higher temperatures and cutoff voltages. Here, we decipher the interphase instability in LHCE-based LMBs with a $Ni_{0.8}Co_{0.1}Mn_{0.1}O_2$ cathode at elevated temperatures. Our findings reveal that the generation of fluorine radicals in the electrolyte induces the solvent decomposition and consequent chain reactions, thereby reconstructing the cathode electrolyte interphase (CEI) and degrading battery cyclability. As further evidenced, introducing an acid scavenger of dimethoxydimethylsilane (DODSi) significantly boosts CEI stability with suppressed microcracking. A $Ni_{0.8}Co_{0.1}Mn_{0.1}O_2$||Li cell with this DODSi-functionalized LHCE achieves an unprecedented capacity retention of 93.0% after 100 cycles at 80 °C. This research provides insights into electrolyte engineering for practical LMBs with high safety under extreme temperatures.

## Introduction

With the increasing demand for energy storage, lithium-ion batteries (LIBs) no longer meet the practical needs for their comparatively low energy density (< 300 Wh $kg^{-1}$).[1] Lithium metal, recognized for its remarkable specific capacity (3860 mAh $g^{-1}$) and low potential (−3.04 V), is pivotal in the forthcoming high-energy-density battery systems.[2] To optimize the energy density of lithium metal batteries (LMBs), the best strategy is to couple the Li metal anode with a high-specific energy cathode. When combined with nickel (Ni)-rich cathodes, like $LiNi_{0.8}Co_{0.1}Mn_{0.1}O_2$ (NCM811), the energy density of LMBs is expected to 500 Wh $kg^{-1}$.[3] Ether-based solvents are highly compatible with Li metal owing to their outstanding reduction stability, yet they are limited by poor oxidation potential (< 4.0 V vs. $Li^+$/Li).[4] Employing high-concentration electrolytes (HCEs)[5] and localized high-concentration electrolytes (LHCEs)[6] has not only improved the reversibility of Li metal anodes but also expanded the operational voltage range of ether-based solvents. This advancement will bring the practical implementation of high-energy-density LMBs.

LHCEs comprise a conductive salt, a solvating (or ionizing) solvent, and a non-solvating (or non-ionizing) diluent.[7] While they retain the solvation structure of HCEs, LHCEs optimize viscosity and wettability,[8] making them among the most advanced electrolytes for NCM811||Li batteries.[9] Nevertheless, the performance of LHCEs at elevated temperatures remains understudied. Thermodynamically, increased temperatures elevate electron energy levels in both the cathode and anode, altering and narrowing the gap between the highest occupied molecular orbit (HOMO) and the lowest unoccupied molecular orbit (LUMO) of electrolytes. This undermines the stability of both the SEI (solid electrolyte interphase) and CEI (cathode electrolyte interphase). Thus, the electrolytes workable at ambient temperatures might falter at elevated temperatures.[10] Practically, broadening the operational temperature range of electrolytes offers advantages in LMBs for varied and extreme environments, such as electric vehicles, space missions, and energy storage systems.[11] Additionally, higher temperatures can initiate parasitic reactions in LMBs, such as SEI/CEI degradation and electrolyte decomposition, leading to significant heat production and potential thermal runaway.[12] Therefore, expanding the operating temperature of electrolytes may not only enhance the tolerance of LMBs to thermal shocks but also boost their safety.[13] Hence, the workable temperature range of electrolytes has recently attracted much attention, which is a crucial normative for assessing their practicality in LMBs, especially at elevated temperatures.[14]

Recently, several strategies for regulating the solvation structure of LHCEs have been reported to suppress the electrode-electrolyte interactions at elevated temperatures. For example, tuning the terminal alkyl chain of ethers in LHCEs could promote the generation of anion-derived SEI/CEI species.[15] Fluorinated solvents (e.g., methyl difluoroacetate) are also beneficial to broaden the operating temperature for their wider electrochemical stability window, and enhance thermal stability.[14b,16] Nevertheless, their performance in NCM811||Li batteries is unsatisfactory. These challenges arise from the need to simultaneously address the thermodynamic interface instability against the Li anode and the NCM811 cathode.[17] Moreover, previous research usually relies on a trial-and-error method due to the lack of profound understanding of the intrinsic design principle. It is essential to elucidate the correlation between molecular structures in the electrolyte and battery capacity degradation, aiming to guide the systematic design of high-performance electrolytes with a broad temperature range.

In this work, taking a NCM811||Li cell coupled with a typical ether-based LHCE composed of lithium bis(fluorosulfonyl)imide (LFSI), 1,2-dimethoxyethane (DME), 1,1,2,2-tetrafluoroethyl-2,2,3,3-tetrafluoropropyl ether (TTE) as a representative, we have explored the interphase stability of the LHCE at 80 °C and unveiled the related degradation mechanism. It is revealed that the capacity decay originates intrinsically from the NCM811 cathode side at high cutoff voltages and elevated temperatures, rather than the Li anode side. Fluorine radicals from $FSI^-$ anions attack an α-H atom in ethers, producing HF and defluorinated $N(SO_2)_2^{3-}$ species. Those by-products are unstable and undergo chain reactions on the surface of the NCM811 cathode, leading to the continuous reconstruction and microcracking of the cathode electrolyte Interphase (CEI). As a result, the capacity of the NCM811||Li battery decays rapidly when operating at 80 °C. To further elucidate the radical-attacking failure mechanism, we propose an acid-scavenger strategy for stabilizing the LHCE and suppressing the CEI degradation at elevated temperatures.[18] As expected, a dimethoxydimethylsilane (DODSi)-functionalized LHCE enables the NCM811||Li cell with an unprecedented capacity retention of 93.0% after 100 cycles at 80 °C, and 81.8% of the initial capacity is retained even after 200 cycles, outperforming conventional LHCEs in terms of operating temperature ranges. This work sheds light on the interphase instability of LHCE-based LMBs at elevated temperatures.

## Results and Discussion

To understand the electrochemical behavior of the LHCE (LiFSI : DME : TTE = 1 : 1.2 : 3 in molar ratio) at different temperatures, the NCM811||Li batteries were assembled using the LHCE electrolyte, and galvanostatic charge-discharge (GCD) tests were implemented at different temperatures. The capacity retention of the batteries over 100 cycles at 28 °C, 60 °C, 70 °C, and 80 °C is 99.4%, 93.0%, 88.9%, and 72.2%, respectively (Figure 1a). However, the battery was inoperative at 90 °C. Therefore, the operating temperature boundary for the NCM811 cathode in LHCE was defined as 80 °C. The capacity decreased rapidly (to nearly 0 mAh $g^{-1}$) in 200 cycles (Figure S1).

The NCM811 cathodes at 80 °C exhibited electrochemical profiles almost identical to those at 28 °C (Figures S2 and S3),

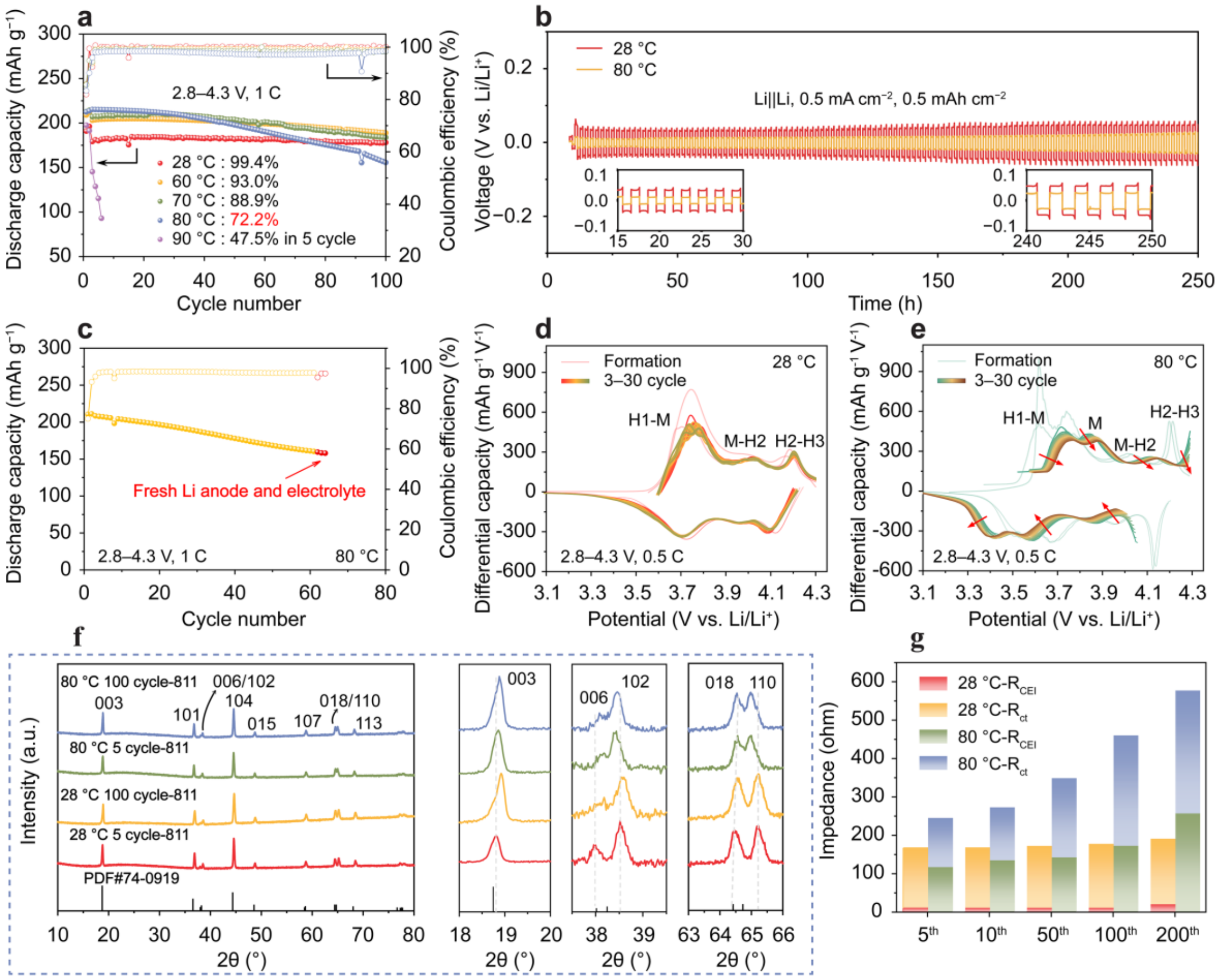

*Figure 1.* Identification of the failure causes of NCM811|LHCE|Li batteries at 80 °C. a) Cycling stability of NCM811|LHCE|Li batteries at different temperatures. b) Cycling stability of Li/Li symmetrical cells with LHCE at temperatures of 28 °C and 80 °C. c) Cycling performances of NCM811|LHCE|Li cells after the replacement of lithium anode at 80 °C. The differential capacity (*dQ/dV*) curves of NCM811|LHCE|Li cells at d) 28 °C and e) 80 °C. f) XRD patterns of the NCM811 cathodes upon different cycles at 28 °C and 80 °C. g) Impedance spectra of the NCM811|LHCE|Li cells at different cycles at 28 °C and 80 °C.

revealing that elevated temperature did not substantially affect the lithiation/delithiation behavior.[19] However, their continuously decreased discharged average voltage reveals the hindered reaction kinetics (Figure S4).[20] To distinguish whether the NCM811 cathode or the lithium metal anode has a more significant influence on cycling stability at 80 °C, a Li||Li symmetrical cell was assembled, which exhibits stable cycling performance for over 250 h without a short circuit at a current density of 0.5 mA cm$^{-2}$ and 80 °C (Figure 1b). This result indicates that the Li metal anode maintains stability with the LHCE at 80 °C. Furthermore, we replaced the previously used electrolyte and the cycled lithium anode with fresh ones, subsequent cycling tests of the new NCM811||Li cells were conducted under identical operating conditions as prior tests. As shown in Figure 1c, the specific capacities of the NCM811||Li cell were not been restored after the replacement of the lithium metal anode and the corresponding electrolyte. Hence, in the NCM811||Li cells, the electrochemical stability of the NCM811 cathode should be more accountable for the cycling performance, rather than the lithium anode.

To better understand the degradation behavior on the NCM811 cathode, the differential capacity analysis (*dQ/dV*) was carried out (Figures 1d and 1e). The *dQ*/*dV* curves at 28 °C show little change during the initial 30 cycles. This is due to the LiF-enriched CEI layers formed on the cathode surface, which could inhibit the interphasial side reactions and stabilize the bulk structure of NCM811. In contrast, the overall curves at 80 °C became increasingly polarized (shifts to a higher voltage) and decreased in intensity during cycling, where the polarized curves reflect the hindered electrochemical kinetics. Also, the peak intensities for H1-M, M-H2, and H2-H3 are decreased by 20.7%, 5.0%, and 58.7%, respectively (Table S1). In particular, the irreversible phase transition of H2-H3 becomes more obvious, leading to reduced reversibility of $Li^+$ (de)intercalation. The above *dQ/dV* analysis shows that the degradation of the NCM811 cathode operating at 80 °C not only decreases the amount of $Li^+$ inventory that can be reversibly (de)intercalated, but also affects the kinetic loss of usable $Li^+$ ions. To determine which factor is more crucial for the NCM811 degradation, we supplement the electrochemical cycling of NCM811||Li cells at a slower current rate of 0.1C after 50 cycles at 1C at 80 °C. It is found that the capacity recovers 73.3% of its total capacity loss (Figure S5). Thus, the cycle degradation is more correlated with the kinetic loss of usable $Li^+$ ions. Furthermore, we evaluated the bulk structure of active NCM811 upon cycling. Even after 100 cycles, the NCM811 crystallites still retain their layered structure, as suggested by their X-ray diffraction (XRD) patterns (Figure 1f): there is no obvious shift with respect to the main peaks, and the separation of (006)/(102) and (108)/(110) peaks remains clear.[21] Therefore, the CEI and surface degradation on NCM811 are more likely to be responsible for its capacity loss at 80 °C, rather than the failure of its bulk structure. To further demonstrate this, the evolution of electrochemical impedance spectra (EIS) for the NCM811||Li cells with the LHCE at 80 °C was studied (Figure S6). The high-frequency range semicircle in the EIS curves can be ascribed to the interfacial resistances of the NCM811 cathode, while the second semicircle in the intermediate frequency could be attributed to charge transfer resistance at the cathode surface.[22] As shown in Figure 1g, the rapidly increasing interphasial resistance ($R_{CEI}$) and charge transfer resistance ($R_{ct}$) of NCM811 at 80 °C compared with the case at 28 °C manifest a continuously growing CEI film and surface degradation.

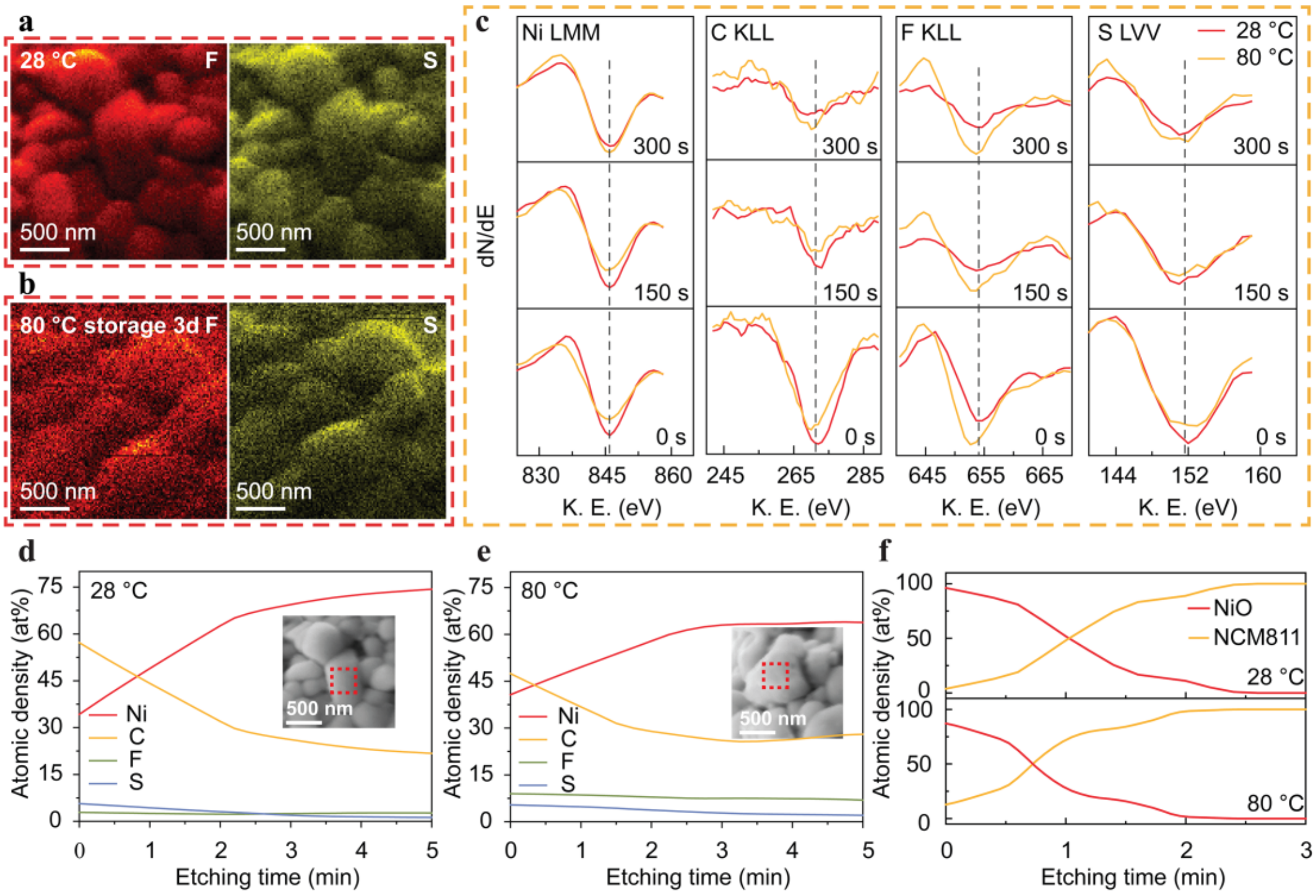

***Figure 2.*** Thermal stability of the cycled NCM811 cathode by AES analysis. a) High-resolution depth analysis in the AES spectra of Ni, C, F, and S by $Ar^+$-beam etching before and after heating at 80 °C. Elemental mapping images of F and S by Auger electron b) at 28 °C and c) heating at 80 °C. The atomic ratios of CEI through $Ar^+$ depth profiling d) at 28 °C and e) heating at 80 °C. f) Auger electron peaking for the Ni element by target factor analysis.

Then we investigated the failure behavior of the CEI. Intrinsic thermal stability is a precondition for the stable operation of Ni-rich cathodes at elevated temperatures. Prior to investigating the degradation behavior of the CEI at 80 °C, we employed Auger electron spectroscopy (AES) to confirm the intrinsic thermal stability of the CEI. The merits of AES, including high spatial resolution and high sensitivity to ionization of core level and outer shell energies, enable it to provide fingerprint information about the surface chemical bonding states of individual particles.[23] This information is used to gain insight into the chemical stability of the CEI and cathode surface at 80 °C. Auger spectra usually display as differentials (*dN/dE*) to enhance the sharp Auger peaks and deemphasize the relatively intense background. On the 50 cycled NCM811 surface, the elements of S (LVV, 151.8 eV), F (KLL, 654.1 eV), C (KLL, 271.6 eV), Ni (LMM, 845.9 eV), and O (KLL, 509.7 eV) were detected by the AES survey spectrum (Figure S7a). The F and S elements correspond to LiF and $S–O_x$, respectively, while C is identified as a series of decomposition products of the solvent, such as $Li_2CO_3$.

After storage in the $Ar_2$ atmosphere at 80 °C for 3 days, no significant variations in energy intensity and *dN/dE* peak shift were detected (Figure S7b), suggesting a stable chemical environment for the elements in the CEI at 80 °C. This is further confirmed by the homogeneous elemental distribution of S and F after storage at 80 °C (Figures 2a, b). The stability of the CEI along the depth direction was further explored by depth profiles of the 50-cycled NCM811 using $Ar^+$-beam etching, and the AES spectra of C, F, S, and Ni are shown in Figure 2c. It is worth noting that there are no significant shifts in the peak positions of these elements even after etching for different durations.

The atomic ratios of the CEI through $Ar^+$ depth profiling are illustrated in Figures 2d and 2e, indicating that at 80 °C the chemical environment of the constituents remains stable at different depths within the CEI. A target factor analysis is used to determine the distribution of different elemental chemical states across these depths.[24] Interestingly, two components were detected in the peak for the element Ni during the etching process, which could be identified as Ni in the NCM811 layered phase and NiO rock-salt phase, respectively. The rock-salt phase usually arises from the corrosion of the NCM811 surface by the electrolyte during cycling and will hinder the reaction kinetics.[25] As shown in Figure 2f, the proportion of NiO drops to zero after 2-min etching. This trend remains unchanged even after storage at 80 °C, implying that the NCM811 cathode possesses intrinsic thermal stability and does not tend to form the rock-salt phase during storage at this temperature. Therefore, both the CEI and the NCM811 cathode material demonstrate inherent thermal stability at 80 °C.

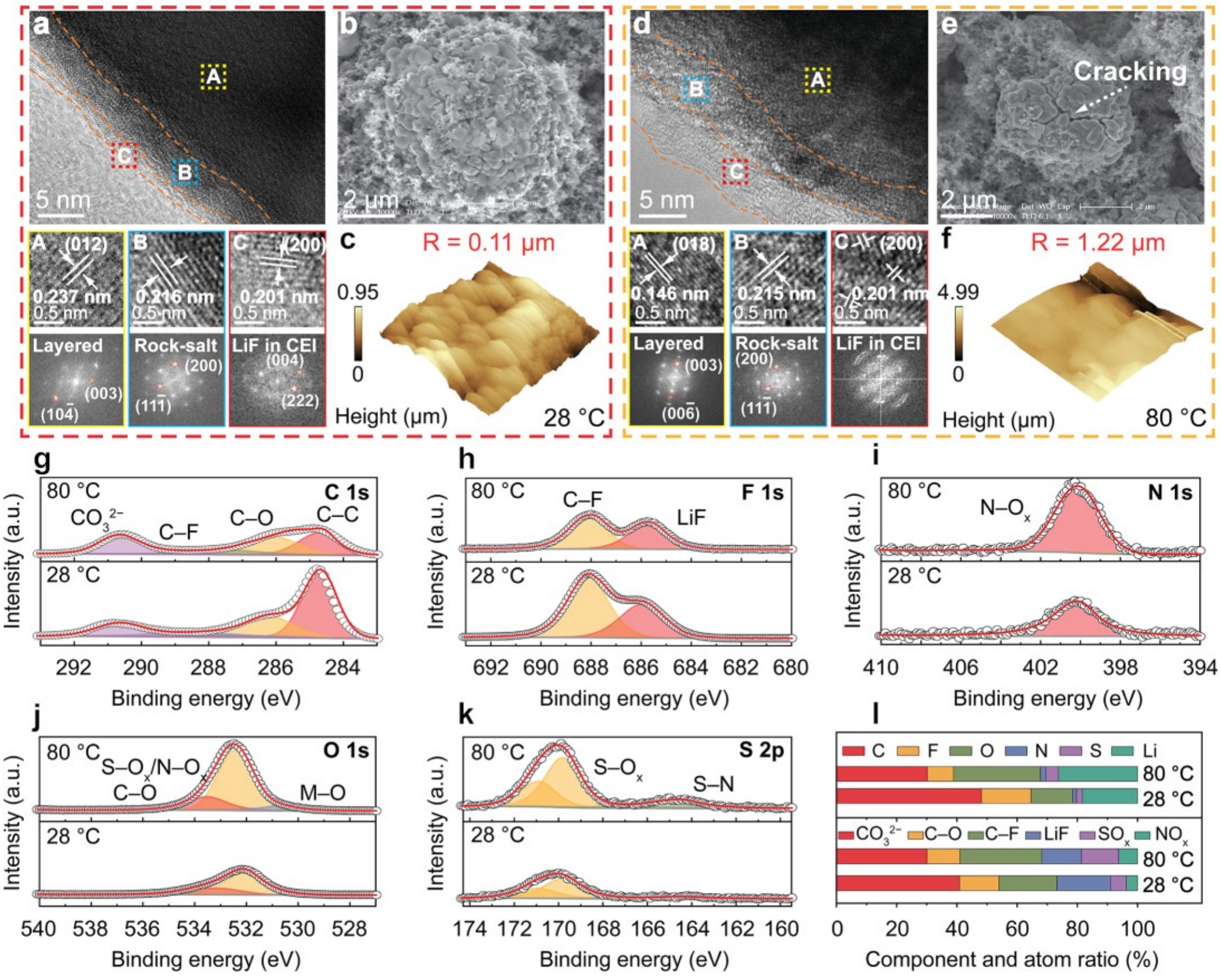

*Figure 3.* HRTEM and XPS analyses of the CEI layers on the surface of the NCM811 cathode at 80 °C. HRTEM images and corresponding Fast Fourier Transformation (FFT) images for the cycled NCM811 particles operating with LHCE at a) 28 °C and b) 80 °C. Corresponding scanning electron microscopy (SEM) images at c) 28 °C and d) 80 °C. AFM images at e) 28 °C and f) 80 °C. XPS results for the NCM811 electrodes after 50 cycles at different temperatures: g) C 1s, h) F 1s, i) N 1s, j) O 1s, and k) S 2p. l) Corresponding element and component ratios of species in the CEI layers.

Then high-resolution transmission electron microscopy (HRTEM) was used to probe the degradation behavior of the CEI under operation at 80 °C. As shown in Figure 3a, the layered structure was indexed as (003) and (10$\bar{4}$) for R-3m, the rock-salt phase was indexed as (200) and (11$\bar{1}$) for Fm-3m, and the LiF phase was indexed as (004) and (222). After 50 cycles in the LHCE at 28 °C, a dense and uniform CEI layer of ~2 nm thickness was observed on the NCM811 surface. The CEI layer exhibits a partially crystalline characteristic. Within this layer, crystalline LiF emerges with a d-spacing of approximately 0.20 nm. Given LiF's wide band gap (13.6 eV) and remarkable oxidative stability (6.4 V vs. $Li/Li^+$),[26] a LiF-enriched CEI can proficiently suppress electrolyte-induced side reactions on the NCM811 surface. Benefiting from the protection provided by a high-quality CEI, the thickness of the NiO rock-salt phase on the surface of the NCM811 cathode is only 4 nm (region B, Figure 3a), which is similar to that of fresh NCM811 (Figure S8). Therefore, the surface of the cycled NCM811 does not show any noticeable cracks (Figure 3b). The roughness of the CEI was also examined by atomic force microscopy (AFM). As depicted in Figure 3c, the CEI at 28 °C exhibits the surface of primary particles similar to that of fresh NCM811, with a roughness of about 0.11 μm (Figure S9). In contrast, after operation at 80 °C, the CEI on the NCM811 cathode becomes thicker and more uneven, and its thickness reaches 4–7 nm (Figure 3d). Although the LiF phase is also detectable in the CEI, the lattice arrangement of the LiF phase has become disordered (Region C, Figure 3d). The disruption in the crystal structure breaks the continuity of the LiF crystal lattice, resulting in the formation of numerous grain boundaries that would potentially hinder the lithium-ion transport. Beneath the CEI layer, the thickness of the NiO phase expands to around 7 nm, a significant increase compared to that at 28 °C. Meanwhile, cracks emerge on the surface of the cycled NCM811 (Figure 3e), leading to an increased surface roughness of 1.22 μm (Figure 3f). These findings demonstrate that the CEI layer fails to protect the NCM 811 surface when operated at 80 °C.

X-ray photoelectron spectroscopy (XPS) analysis was performed to further investigate the evolution of the CEI operating at 80 °C (Figures 3g–l). At 28 °C, the polyether-derived segment (C–O, 286.2 eV, C 1s),[27] C–F species (289.3 eV, C 1s), $Li_2CO_3$ (290.8 eV, C 1s), LiF (686.1 eV, F1s), S–$O_x$ (170.3 eV, S 2p) and N–$O_x$ (400.2 eV, N 1s) species are detected on the surface of NCM811, contributing to 40.9%, 19.2%, 13.2%, 17.9%, 5.2% and 3.7% of the component ratios, respectively, which agrees well with the AES and HRTEM results. The robust CEI layer containing rich LiF and organic species enabled by TTE can effectively suppress the side reaction between NCM811 and the electrolyte. In particular, the components of the CEI at temperatures of 28 °C and 80 °C are quite different. The contents of $Li_2CO_3$ and LiF are reduced to 11.0% and 13.1%, respectively, while the contents of S–$O_x$ and N–$O_x$ are increased to 12.4% and 6.3%. Additionally, a trace amount of transition-metal oxide (MO) was observed in the CEI (530.7 eV, O 1s, Figure 3j), signaling the dissolution of TM-ions. It is concluded that elevated temperatures significantly affect both the composition and structure of the CEI. When operated at 80 °C, the initially dense, thin, and uniform CEI becomes rough and uneven, which not only hinders lithium-ion transport but also compromises the effective shielding of the cathode surface to the electrolyte, bringing about the rapid degradation of the NCM811 surface structure.

Given the direct impact of the solvation structure on the CEI components, we conducted a Raman analysis to reveal temperature-induced variations in the solvation structure of LHCE. In Figure 4a, the Raman peaks appear at 732, 742, and 754 $cm^{-1}$, corresponding to a different form of the FSI anion, presenting as a free ion or solvent-shared ion pair (SIP), a contact ion pair (CIP), and an agglomerate (AGG), respectively.[28] Notably, the AGG and CIP ratios increase (72% vs. 70%) relative to ambient conditions, which is due to the stronger bonding between $Li^+$ and $FSI^-$. Molecular dynamics (MD) simulations further determined the coordination numbers of $Li^+$ across varied temperatures (Figure S10). As temperature rises, the coordination number (CN) of $Li^+$ with $FSI^-$ increases (3.6 *vs.* 3.4), whereas the CN with DME decreases (1.5 *vs.* 1.2), which is consistent with the Raman results (Figures 4b, c). Theoretically, the formation of a LiF-rich CEI on the NCM811 surface becomes more favorable, promoting the stability of the NCM811 cathode. Nevertheless, such findings conflict with the XPS results and the observed capacity degradation in NCM811.

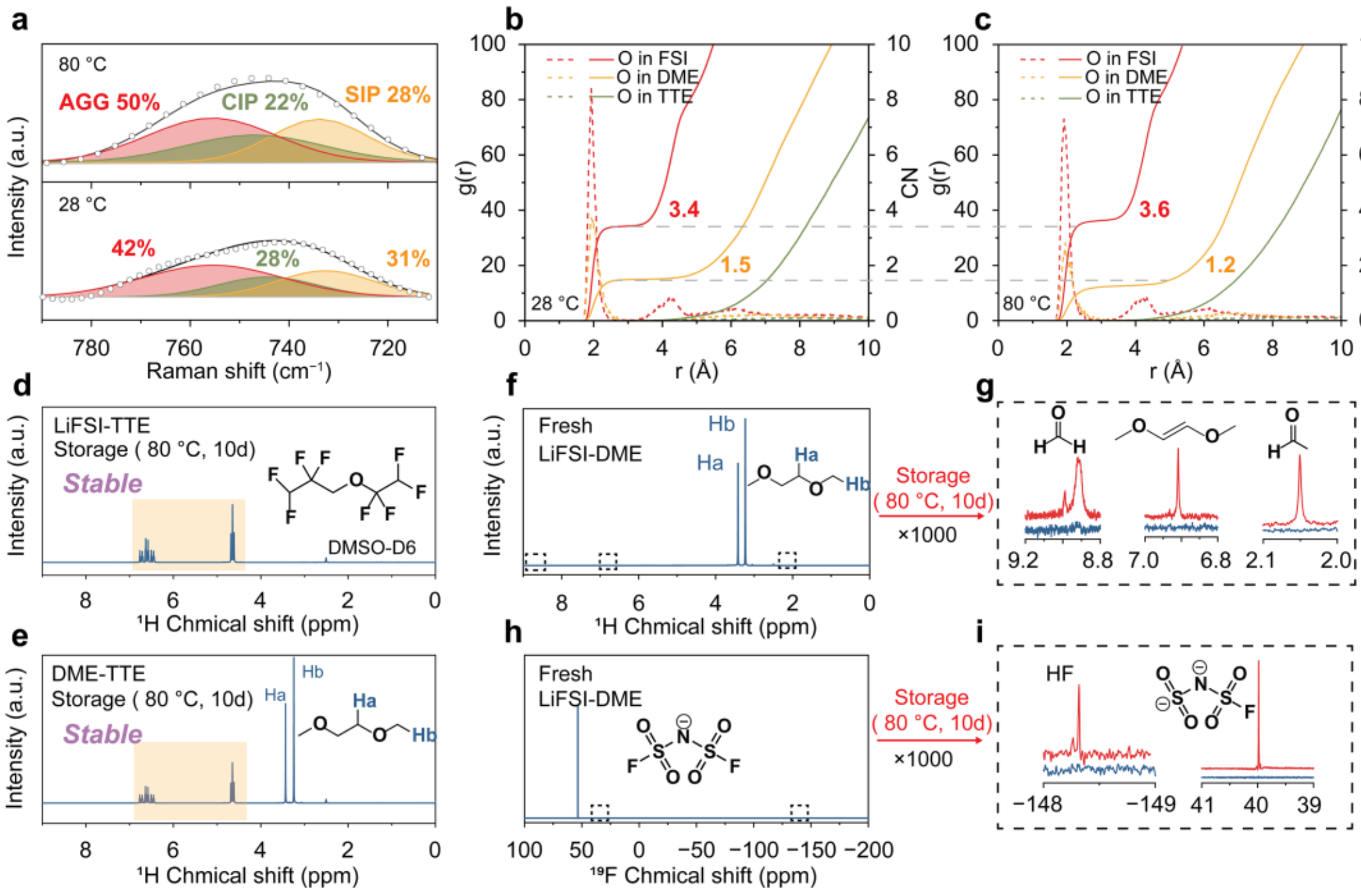

*Figure 4.* Degradation analysis of the LHCE after storage at 80 °C for 10 days. a) Raman spectra of LHCE at 28 °C and 80 °C. The peaks near 732, 742, and 754 $cm^{-1}$ are assigned to SIP, CIP, and AGG. Radial distribution function (RDF) and coordination number (CN) of $Li^+$ in LHCE b) at 28 °C and c) at 80 °C, respectively. $^1H$ NMR spectrum of d) LiFSI-TTE and e) DME-TTE after 80 °C storage for 10 days. f) $^1H$ NMR and g) $^{19}F$ NMR spectrum of fresh LiFSI-DME. Magnified views of regions of the LiFSI-DME before and after 80 °C storage for 10 days. h) $^1H$ NMR and i) $^{19}F$ NMR.

Considering that the elements of S, N, and F in the CEI are all derived from $FSI^-$ anions, and due to the noticeable change in the relative ratios of S, N, and F from XPS results, it is speculated that the reaction path of the electrolyte on the surface of the NCM811 cathode has changed when operating at 80 °C. Moreover, the decrease in the proportion of $Li_2CO_3$ might be attributed to an increase in the acidity of the electrolyte. To verify these speculations, the electrolyte stability was further investigated by combining the components of the LHCE in pairs, including LiFSI-DME, LiFSI-TTE, and DME-TTE, and each pair was stored at 80 °C for 10 days. The relative ratio of different components is kept identical to ensure repeatability. The composition of the resulting solution was analyzed using nuclear magnetic resonance (NMR) spectroscopy to identify potential parasitic reaction products. The $^{19}F$ NMR results show that the peak at 53.2 ppm corresponds to $FSI^-$ (Figure S11a), and the peaks at −90.9, −125.7, and −139.1 ppm are assigned to the fluorine element in TTE (Figure S11b). DME does not generate any signal in $^{19}F$ NMR (Figure S11c). After storage at 80 °C for 10 days, no significant color change was observed in the solutions of LiFSI-TTE and DME-TTE pairs (Figure S12). Additionally, there were no new peaks detected in the $^1H$ NMR and $^{19}F$ NMR spectrum, implying that the LiFSI-TTE and DME-TTE solutions remained stable at 80 °C (Figures 4d and 4e). In the LiFSI-DME pair, small amounts of HCHO (8.92 ppm), $CH_3O(CH)_2OCH_3$ (6.92 ppm), and $CH_3CHO$ (2.05 ppm) were found after storage at 80 °C (Figures 4f and 4g), which signals that the DME underwent oxidation reactions. A characteristic peak for HF appeared at −148.3 ppm (Figures 4h and 4i),[15] indicating that the $FSI^-$ underwent a defluorination reaction. Since defluorination would reduce the steric hindrance of $FSI^-$, the peak of $FSI^-$ in the $^{19}F$ NMR spectrum shifts to the lower field, corresponding to the new peak of $SO_2NSO_2F^{2-}$ at 40.0 ppm (Figure 4i). To further explore the defluorination mechanism of $FSI^-$ and its effect on the CEI, we applied ab-initio molecular dynamics (AIMD) on the LiFSI-

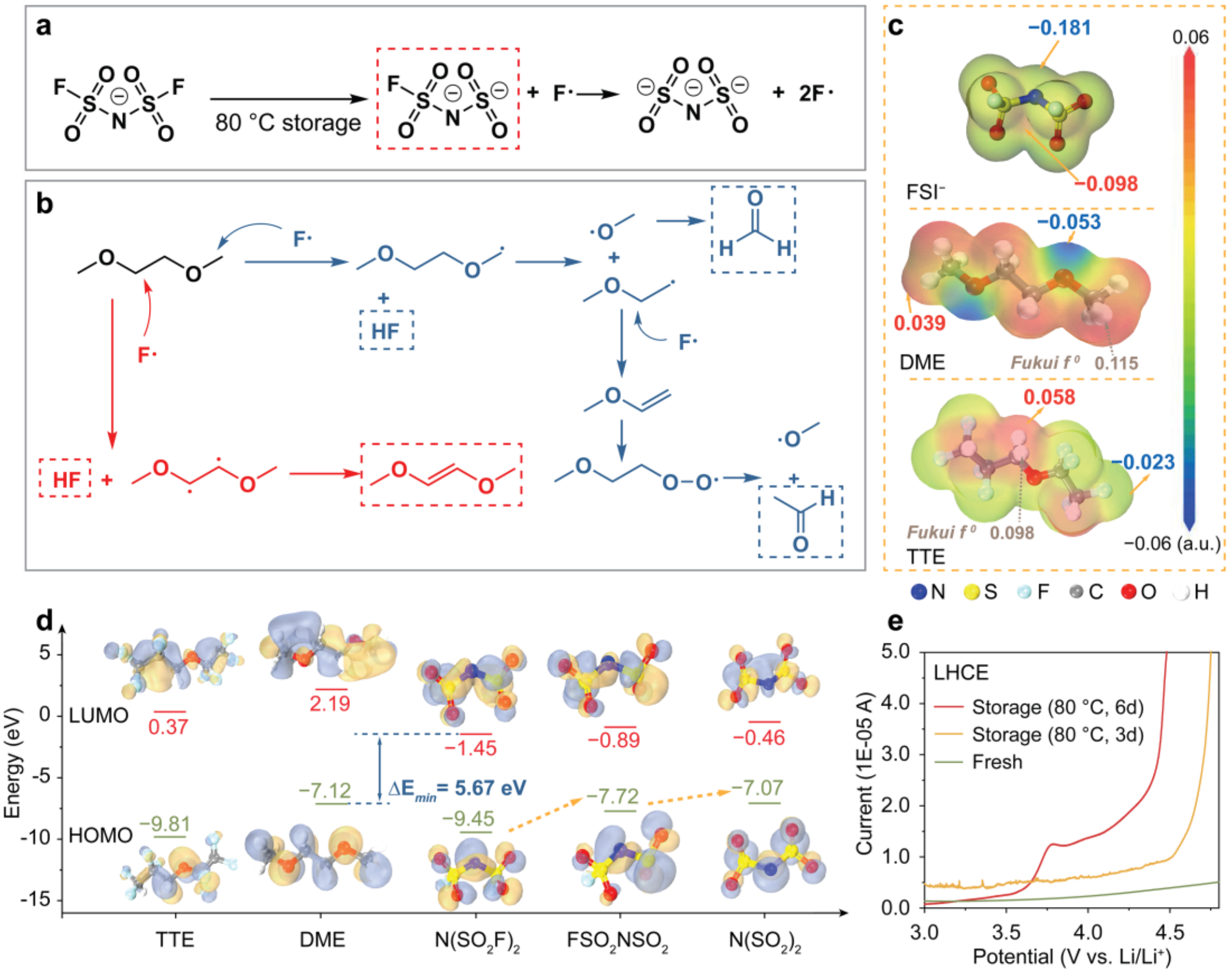

***Figure 5.*** The degradation mechanism in LHCE at 80 °C. a) LiFSI defluorination. b) DME. c) Electrostatic potential mapping about electron distribution for $FSI^-$, DME, and TTE. d) LUMO and HOMO energy of the composition and by-products in LHCE. e) LSV curves of Li|LHCE|Al cells after storage for different days at 80 °C.

DME system (Figure S13). Since the thermal decomposition of the electrolyte at 80 °C may require substantial time, while the actual running time of AIMD is much shorter, we implemented an acceleration strategy by increasing the temperature and conducting the AIMD simulation of LiFSI-DME at 500 K.[29] After 1 ps of dynamic steps, the S–F bond in FSI⁻ breaks, then the S–N bond and the alpha-position C–H bond of DME break after 2 ps.

Based on the analysis of AIMD calculations, the possible reaction mechanism between LiFSI and DME is shown in Figures 5a and 5b. It is well-known that the hydrogen atoms at the α-position (α-H) of the ether group are susceptible to attack by nucleophiles. Given the lowest bond order of 0.78 of the S–F bond in FSI⁻ (Figure S14), high temperatures lead to the generation of F radicals and the associated defluorination by-products, $FSO_2NSO_2^{2-}$/$N(SO_2)_2^{3-}$. These F radicals target the α-H in DME, subsequently producing HF species. Further calculations of the electrostatic potential (Figure 5c) were performed, where the electrostatic potential and Fukui $f^0$ coefficients could serve as an indicator of the reaction site. The α-H site exhibits the highest $f^0$ value of 0.115 in DME, where it is particularly vulnerable to being attacked by F radicals. Conversely, the $f^0$ value for α-H is 0.098 in TTE, possibly attributable to the fluorine atoms' bond to α-C altering the electronic environment around the ether group. Consequently, TTE demonstrates minimal reactivity with FSI⁻ at 80°C. These findings suggest that regulating electron distribution through molecular structure designs can potentially improve the high-temperature stability of LHCE.

The density functional theory (DFT) calculations for the frontier molecular orbital reveal that defluorination by-products have a higher HOMO energy level when compared to the FSI⁻.[30] As shown in Figure 5d, among the components of LHCE, the LUMO of FSI⁻ and the HOMO of DME are the closest ($\Delta E_{min}$ = 5.67 eV), making them the most likely to react with each other. After defluorination for FSI⁻, the HOMO of $FSO_2NSO_2^{2-}$ and $N(SO_2)_2^{3-}$ is up to –7.72 and −7.07 eV, respectively, while that of FSI⁻ ($N(SO_2F)_2^-$ ) is −9.45 eV. Therefore, $N(SO_2)_2^{3-}$ anions are more likely to undergo oxidation on the 811-cathode surface compared with FSI⁻ anions, resulting in an increased amount of S–$O_x$/N–$O_x$ species, accompanied by a decrease in LiF species (XPS analysis). As shown by the linear scanning voltammetry (LSV) results (Figure 5e), the oxidation potential of LHCE decreased with the storage time. After storage for 6 days, the oxidation potential of LHCE is reduced to 3.5 V, while that of fresh LHCE is more than 4.5 V, which is consistent with the DFT calculation results. Due to the HF-induced damage to the CEI, these by-products are primed to undergo oxidative decomposition on the cathode surface preferentially, resulting in an increased proportion of S–$O_x$ and N–$O_x$ species. The S–$O_x$ and N–$O_x$ species have a loose and porous structure (as depicted in Figure S15), and fail to effectively prevent the electrolyte from contacting with the NCM811 cathode surface, thus inducing persistent interphasial side reactions. Ultimately, the irreversible phase transition and dissolution of transition-metal ions occur at the NCM811 cathode surface, which is in good agreement with the TEM and XPS results.

Figure 6 depicts the chain reaction mechanism that triggers the CEI failure at 80 °C. Benefiting from the anion-dominated solvation structure, the LiF-rich CEI was formed on the surface of the NCM811 cathode at 28 °C (Figure 6a). At 80 °C, Initially, the interaction between FSI⁻ and DME in the LHCE produces HF and a variety of defluorination by-products, such as $FSO_2NSO_2^{2-}$ and $N(SO_2)_2^{3-}$. Subsequently, HF corrodes the $Li_2CO_3$ species within the CEI, which induces abundant defects within the LiF phase, thereby inhibiting Li-ion transport. Meanwhile, for the lower oxidation potential, the by-products $N(SO_2)_2^{3-}$ would undergo further oxidative decomposition at the cathode surface preferentially over the FSI⁻ anions, yielding increased S–$O_x$ and N–$O_x$ species. Eventually, the loosely structured S–$O_x$ and N–$O_x$ species cannot prevent the electrolyte from contacting the NCM811 cathode surface, which inevitably and continuously induces interphasial parasitic reactions (Figure 6b). As a result, irreversible phase transitions along with the dissolution of transition-metal ions on the surface are exacerbated, causing a continuous decay in cathode capacity when operated at 80 °C (Figure 6c).

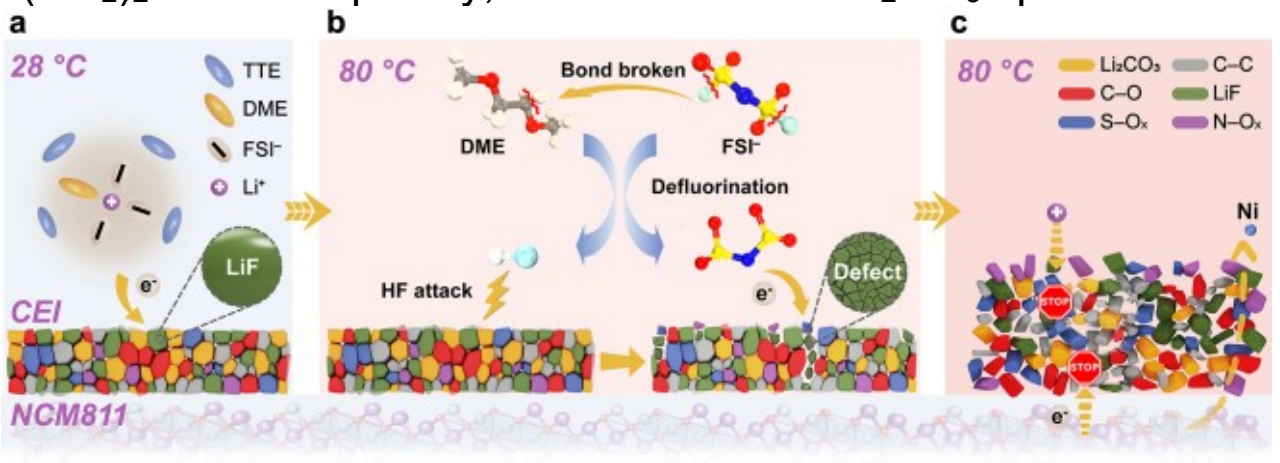

**Figure 6.** Schematic illustration for the degradation mechanism of NCM811|LHCE|Li battery at elevated temperatures. a) The CEI of NCM811|LHCE|Li battery at 28 °C. b) The dynamic evolution for the CEI of NCM811|LHCE|Li battery at 80 °C. c) The CEI of NCM811|LHCE|Li battery at 80 °C after evolution.

To further elucidate the above HF-attacking failure mechanism, we propose an acid-scavenger strategy to address the issue of the unstable CEI at elevated temperatures, whereby a DODSi additive of 0.5 wt% was introduced into the ether-based LHCE. As shown in Figures 7a and S16, the addition of DODSi to the LHCE succeeds in increasing the capacity retention of the NCM811||Li cells from 55.1% to 93.0% over 100 cycles at 80 °C, in comparison to the conventional LHCE. Notably, the HF signal is hardly detectable for the DODSi-functionalized LHCE after storage at 80 °C for 10 days, confirming the excellent acid removal ability of DODSi (Figure 7b). Compared to previous reports on high-temperature electrolyte design, our results exhibit all-around performances to a greater extent, which for the first time represent the NCM811||Li cells stably working at 80 °C in terms of a high mass loading of the active NCM811 of 18 mg $cm^{-2}$ (Figure 7c). It is demonstrated that DODSi can not only bind with $H^+$ by acid-base interaction and thereafter, but also scavenge separated fluorine radicals due to the high binding affinity of Si toward F (Figure 7d).[18b] Hence, the CEI on the NCM811 cathode surface remains stable at 80 °C (Figure 7e), thereby significantly improving the performance of the NCM811||Li cells operating at elevated temperatures.

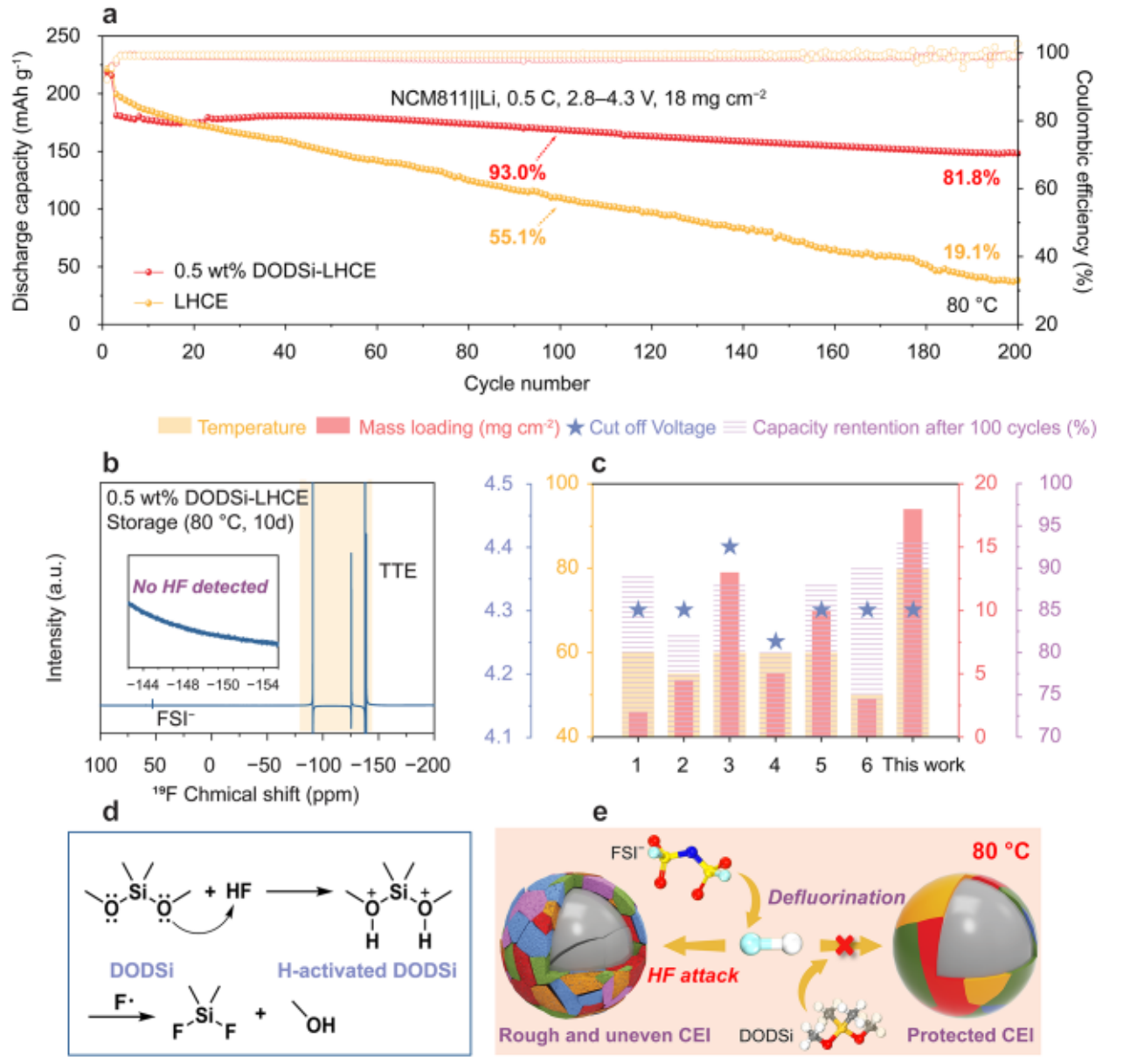

**Figure 7.** Performance of NCM811||Li batteries with the DODSi-functionalized LHCE. a) Cycling stability of NCM811|LHCE|Li batteries at 80 °C. b) $^{19}$F NMR spectrum of 5 wt% DODSi-containing LHCE storage at 80 °C for 10 days. c) Comparison of recently reported NCM811||Li batteries at extreme temperatures. (1–6 are cited in [14b, 15, 31] respectively.) d) Mechanism of DODSi for HF scavenging reaction. e) Schematic illustration for the mechanism of DODSi in LHCE.

## Conclusion

In summary, we have deciphered the interphase instability of the ether-based LHCE at high temperatures and shed light on the relationship between the electronic environment in the LHCE and NCM811 degradation. Our findings demonstrate fluorine radicals in $FSI^-$ targeting the α-H in ethers, subsequently yielding HF and the defluorinated $N(SO_2)_2^{3-}$ species. HF-induced the decomposition of $Li_2CO_3$ species, and the preferential oxidation of defluorinated $N(SO_2)_2^{3-}$ species occurs simultaneously, intensifying irreversible phase transitions on the NCM811 surface and increasing interfacial impedance. Suppressing the intermediate steps of the chain reaction is an effective way to boost the high-temperature electrochemical performance of the LHCE toward practical high-energy-density LMBs. As evidenced, introducing an acid-scavenger of DODSi into the LHCE has realized stable cycling of NCM811||Li cell with a high capacity retention of 93.0% after 100 cycles at 80 °C, surpassing conventional LHCEs operating at elevated temperatures. This work provides valuable insights into electrolyte and interphase engineering to extend the operating range of LHCEs and Ni-rich cathodes toward future practical LMBs in terms of harsher environments at high temperatures.

## Acknowledgements

This work is supported by the National Natural Science Foundation of China (Nos. 52272206 and 51972132). The authors also thank the Analytical and Testing Center of HUST for XRD, Raman, SEM, AES, and TEM measurements.

## Conflict of Interest

The authors declare no competing interests.

## Data Availability Statement

The data that support the findings of this study are available from the corresponding author upon reasonable request.

**Keywords:** NCM811-Li batteries • localized high-concentration electrolyte • elevated temperatures • electrode-electrolyte interphase • degradation mechanism

## Entry for the Table of Contents

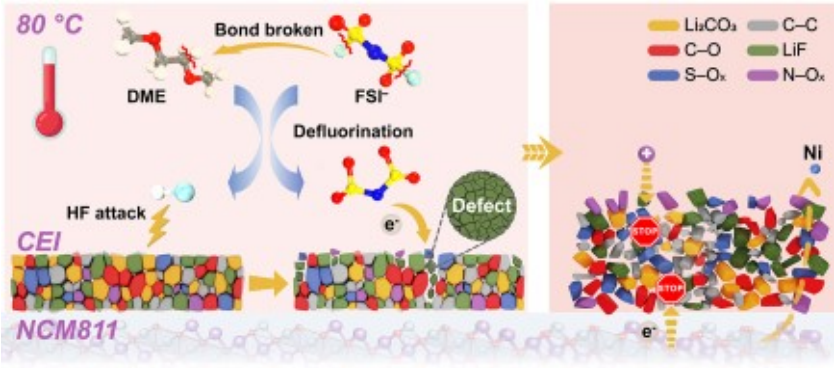

In this study, we decipher the interphase instability at elevated temperatures for the NCM811||LHCE|Li battery. The fluorine radicals trigger the decomposition of ether molecules and reconstruction of the cathode electrolyte interphase (CEI), further deteriorating high-temperature cycling stability.